\documentstyle[11pt,newpasp,twoside,epsf]{article}
\def\edcomment#1{\iffalse\marginpar{\raggedright\sl#1\/}\else\relax\fi}
\reversemarginpar

\begin{document}
\title{X-Ray Plasma Diagnostics of Stellar Winds in Very Young Massive Stars}
\author{Norbert S. Schulz, Claude R. Canizares, David P. Huenemoerder, Julia C. Lee, Kevin Tibetts}
\affil{Center for Space Research, Massachusetts Institute of Technology,
Cambridge, MA 02139.}

\begin{abstract}
High resolution X-ray spectra of very young massive stars opened a new chapter
in the diagnostics and understanding of the properties of stellar wind plasmas.
Observations of several very young early type stars in the Orion Trapezium demonstrated
that the conventional model of shock heated plasmas in stellar winds is not
sufficient to explain the observed X-ray spectra. Detailed X-ray line diagnostics
revealed extreme temperatures in some of the candidates as well as evidence
for high plasma densities. It is also evident from high resolution spectra
of more conventional early type stars, that not all show such extreme characteristics.
However, the fact that some of the stars show hot and dense components and some do not requires
more understanding of the physical processes involved in stellar wind emissions.
The Orion Trapezium stars distinguish themselves from all the others by their extreme youth.
By comparing the diverse spectral properties of $\theta^1$ Ori A and $\theta^1$ Ori E with those of
$\theta^1$ Ori C, we further demonstrate that X-ray spectral properties of very young
massive stars are far from understood. 
\end{abstract}

\section{Introduction}

Detailed properties of hot winds in massive O- and very early B-type stars so far have been predominantly
studied in spectra at optical and ultraviolet wavelengths, because here thousands of spectral lines
provide a wealth of parameters, which are needed  to test models describing the origin and
propagation of the winds. High resolution UV spectra of massive O-stars show the presence of strong 
resonance lines that exibit P-Cygni profiles as well as an abundance of weaker absorption
lines (see Kudritzki 1992). They are sensitive indicators of mass loss rates as small
as $10^{-7}$ to $10^{-10} M_{\odot}$ yr$^{-1}$, where radio and infrared observations
are insensitive. These developments  now enable theorists to apply
hydrodynamic stationary (Kudritzki 1984, Pauldrach 1987 and refs. therein) as well as non-stationary (NLTE)
radiation driven wind models (Pauldrach et al. 1994, Lamers et al. 1999) to these UV spectra.

In contrast to the development in the UV domain, interpretations of X-ray spectra from stellar winds
remained quite modest since their first observations with {\sl Einstein } (Seward et al. 1979). The main
reason is that the theory of line radiation-driven stellar winds explains 
the mass loss from a hot luminous star in the UV (Pauldrach et al. 1994, Lamers et al. 1999), 
but doesn't predict observed X-ray fluxes.
To date no definitive model for the production of X-rays in stellar winds exists, although
models involving shocks forming from instabilities within a radiatively driven wind (Lucy $\&$ White 1980,
Owocki, Castor, $\&$ Rybicki 1988, MacFarlane $\&$ Cassinelli 1989, Feldmeier et al. 1995) seemed 
most successful so far reproducing the observed X-ray spectra of kT $\sim$ 0.5 keV temperature.

The biggest obstacle in analyzing X-ray spectra was the modest spectral resolution of detection devices used in 
previous X-ray missions, which did not to resolve  emission and absorption lines. 
In that respect the main results from {\sl Einstein} and {\sl ROSAT } were based on global properties from a large sample
of O- and B-type stars (Chlebowsky et al. 1989, Bergh\"ofer et al. 1996). 
Among these results were that the X-ray luminosity scales with
the bolometric luminosity with a typical median ratio of log (L$_x$/L$_{bol}$) $\sim$ -6 to -7. 
{\sl ROSAT} with its mediocre energy resolution
but enhanced soft sensitivity was probably the first good instrument to investigate soft X-ray spectra
in the 10 to 80~\AA~ region close to the far UV domain. In a first major step forward, Hillier et al. (1993)
fitted well exposed spectra of $\zeta$ Pup with detailed NLTE models under the assumption that
the X-rays arise from shocks distributed throughout the wind and allowing for recombination in the outer
regions of the stellar wind. The best fits predicted two temperatures of log T(K) of 6.2 to 6.7 with shock velocities
around 500 km s$^{-1}$. Based on this approach Feldmeier et al. (1997) assumed that the X-rays originate
from adiabatively expanding cooling zones behind shock fronts, which are decribed by a post shock temperature
and a volume filling factor. These results were also compared to results from Cohen et al. (1996), who used
{\sl ROSAT} and {\sl EUVE} data to constrain high-temperature emission models in the analysis of the B-giant $\eta$CMa.
Here it turned out that a continuous temperature distribution is needed rather than a single or even two
temperature models. Clearly the result of that comparison remained inconclusive, since both views appear
rather plausible but indistinguishable in the spectra.

Besides the need for more than one temperature models, {\sl ROSAT } spectra offered a limited bandpass only and 
most of the wind spectra were found to peak around 0.8 keV. Here CCD spectrometers, like the ones onboard {\sl ASCA }
and {\sl Chandra}, provide a medium energy resolution throughout a band pass of 0.1 to 10.0 keV. 
Several observations of massive early type
stars, i.e. $\tau$ Sco (Cohen et al. 1997) or $\eta$ Car (Tsuboi et al. 1997), indicated that there is a hard
component to some of these spectra of 2 keV and higher. Although these two examples may resemble quite extreme cases
of massive stars, observations of the $\rho$ Ophiuchi Cloud (Kamata et al. 1997), IC 348 (Nagase et al. 1999),
and the Orion Trapezium (Yamauchi et al. 1996, Schulz et al. 2001) indicated the existence of a similar hard component
in the spectra. The Orion Trapezium stars in this respect spark special interest, because here it was shown
in an ensemble of massive early type stars, that some of their spectra indeed require a hard spectral component
of up to kT = 6 keV, some do not (Schulz et al. 2001).    

\section{First Results from High Resolution Spectra}

With the launches of {\sl Chandra} (High Energy Transmission Grating Spectrometer HETGS, Canizares et al., in preparation)
and {\sl XMM} (Reflection Grating Spectrometer RGS, den Herder et al. 2001) in 1999 we now have 
high resolution X-ray grating spectrometers at hand that allow us to finally resolve the 
line emission in X-ray spectra of early type stars. The two first spectra published already showed
an obvious discrepancy in X-ray properties. While the HETGS spectrum of $\theta^1$Ori C (O7 V pne) showed line emission
up to 1.7~\AA ($\sim$ 7 keV) and relatively narrow, symmetric, and unshifted lines (Schulz et al. 2000), the RGS spectrum
of $\zeta$ Pup (O 4 If) indicated relatively broad, asymmetric, and blue-shifted lines at moderate temperatures (Kahn et al. 2001).
So far none of the spectra observed so far, with the exception of $\zeta$ Pup (Kahn et al. 2001, Cassinelli et al. 2001),
exhibit dominating X-ray properties as expected from instability-driven shocks distributed throughout the wind (Waldron and
Cassinelli 2001, Schulz et al. 2000, Gagne et al. 2001).     
In $\zeta$ Ori Waldron and Cassinelli (2001) interprete the lack of a substantial amount of flux in the forbidden lines of
the He-like triplets as an indication that the X-ray emitting zones around $\zeta$ Ori may lie less than
1 R$_{\star}$ from the photosphere, since the large UV radiation field here would destroy the $^3{\rm S}_1 - ^3{\rm P}$ 
transition (Kahn et al. 2001). In fact such an effect is seen in all spectra observed so far and 
reflects the need to account for UV photo-excitation in these sources. In this respect, the interpretation of the
f/i ratio (f = forbidden line flux, i = intercombination line flux) in the He-like triplets in OB stars is 
much more complex and 
requires assumptions about other properties of the X-ray emission regions. Gagne et al. (2001) suggested that 
in less luminous OB stars the line profiles are symmetric and narrower and resemble emission from  
magnetically confined wind shocks. 

Clearly, before can attempt to classify OB stars based on their X-ray line 
properties, we need more high resolution observations. The Orion Trapezium has an age of
$\sim$ 0.3 Myr (Hillenbrandt 1997) and its early type members represent a canonical sample
to study the X-ray properties of very young massive stars. In this article we compare the properties
of Trapezium members $\theta^1$Ori A and E with the ones from the brightest member $\theta^1$Ori C.

\section{$\theta^1$Ori C}

Schulz et al. (2000) already deduced some remarkable properties from the first X-ray spectrum. Depite the fact
that the lines appear symmetric and show no significant blue-shift, line flux ratios from He-like and Li-like
iron indicate temperatures of $\sim$ 60 Million K and a study of H- and He-like lines ratios from lower Z ions
indicate that we see a whole range of temperatures down to 5 Million K. The line widths appear relatively
narrow with Doppler velocities between 200 and 800 km s$^{-1}$. In Figure 1 (top) we show an  
HETGS spectrum of $\theta^1$Ori C, which was compiled from all the data presented in Schulz et al. (2000) plus
all the data from an observation taken three weeks later. The total exposure time is 85 ks. The spectrum is  corrected
for interstellar absorption, where we assumed the column density derived by Schulz et al. (2001)
for the Orion Trapezium stars. For the spectral modelling we apply a collisionally ionized plasma
as calculated by the Astrophysical Plasma Emission Code (APEC, Smith et al. 2001), 
which is implemented in the Interactive Spectral Interpretation
Software (ISIS, Houck \& DeNicola 2000). To account for the wide temperature
range we apply several temperature components, thermal line profiles, and Doppler broadenings between
200 and 800 km/s. 
A preliminary fit of such a model to the spectrum is shown in the bottom part of Figure 1. It includes
five temperatures between 6$\times10^5$ K and 6$\times10^7$ K with dominating contributions from two
temperature components above $10^7$ K.
Although not all details are worked out yet, the model provides already a very good representation of the spectrum.
The model also does not require significant deviations from a 
solar abundance distribution and we identify all major Fe lines expected for Fe XVII to Fe XXIV.

\section{$\theta^1$Ori A and $\theta^1$Ori E}

$\theta^1$Ori A and $\theta^1$Ori E are the second and third brightest X-ray sources in the Orion Trapezium.
Although
assumed to be of very early type as well, the optical properties of $\theta^1$Ori A and $\theta^1$Ori E are not very well 
documented (see Schulz et al. 2001 and refs. therein). Although they appear only 
four ($\theta^1$Ori E) and eight ($\theta^1$Ori A) times less luminous in X-rays compared to $\theta^1$Ori C, which 
is still in the range expected for late O- and early B-type stars (Bergh\"ofer et al. 1996),
their optical fluxes relative to $\theta^1$Ori C are much smaller than these factors indicate (Petr et al. 1998).
In the case of $\theta^1$Ori E this is most extreme indicating that log (L$_x$/L$_{bol}$) is well above -6, which
is untypical for O-type stars. 

\begin{figure*}
\plotone{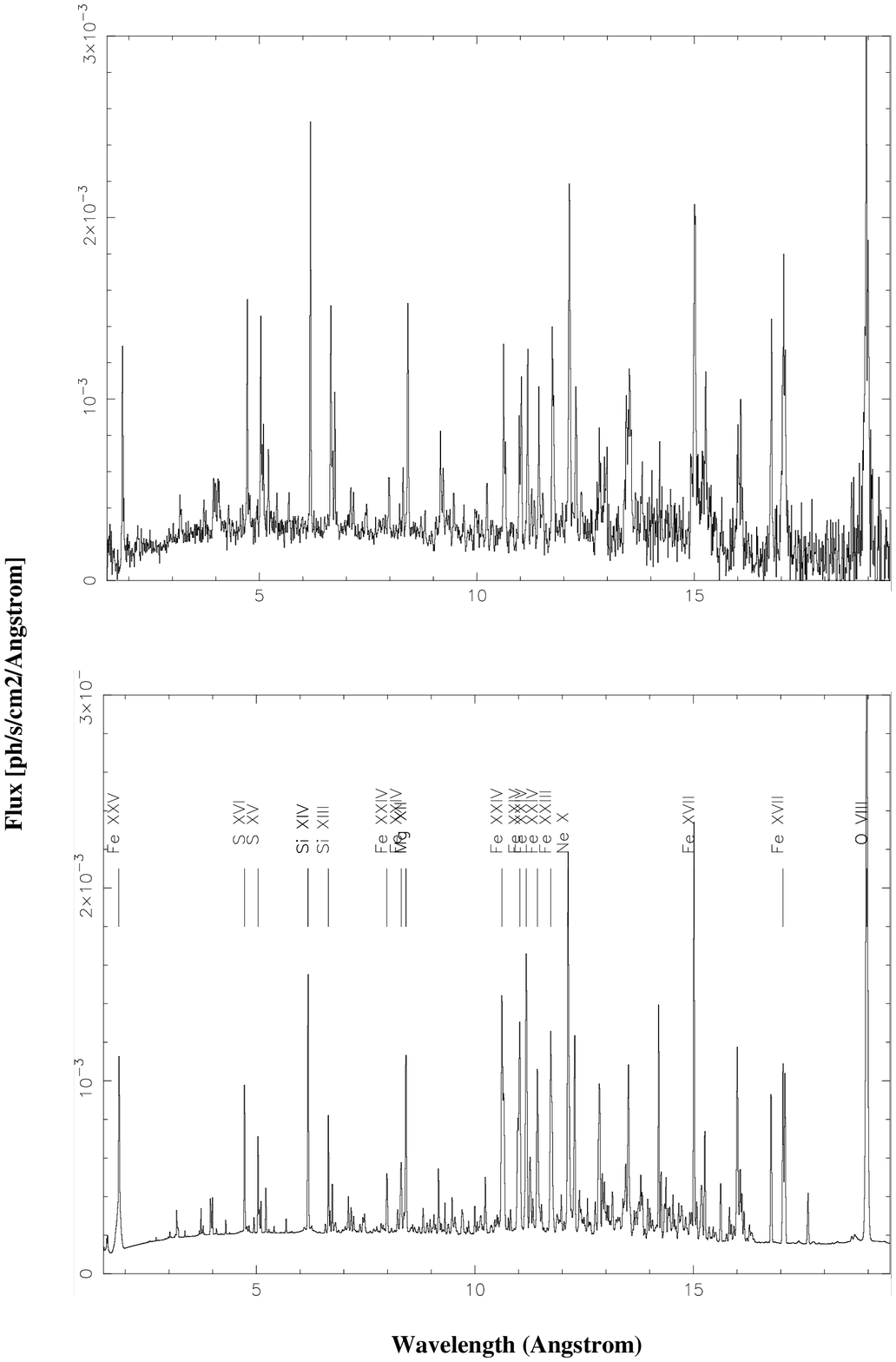}
\caption{The HETGS spectrum of $\theta^1$ Ori C for an exposure of 83 ks (top). A multi-temperature
APED model of $\theta^1$ Ori C assuming a collisionally ionized plasma and solar abundances.}
\end{figure*}

\begin{figure*}
\plotone{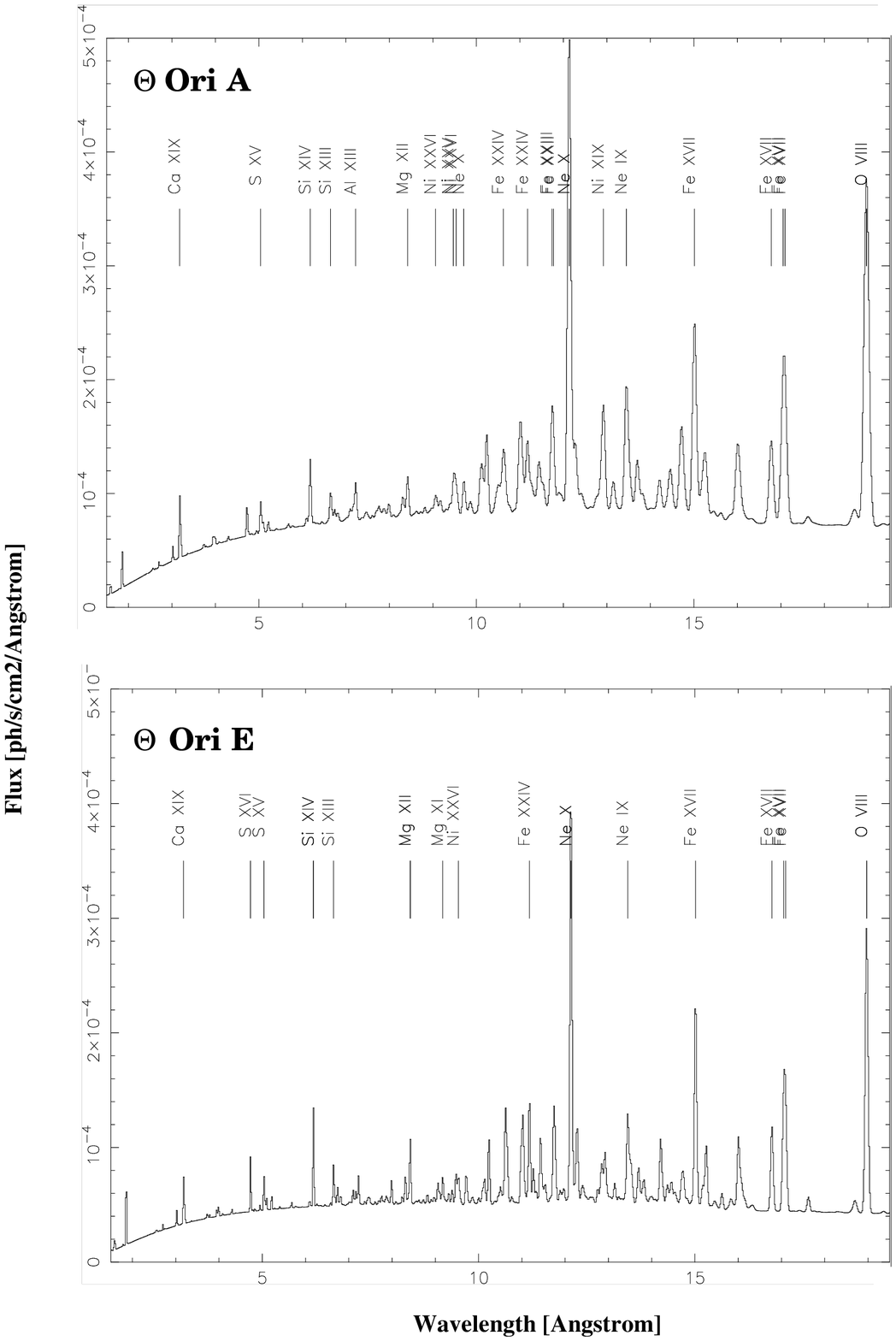}
\caption{Multi-temperature models assuming a collisionally ionized plasma as calculated with APED fitted for
$\theta^1$ Ori A (top) and $\theta^1$ Ori E (bottom). The model assumes reduced abundances compared to solar
for elements higher than Z=11 and lower then Z=27}
\end{figure*}

We applied the multi-temperature model as applied to $\theta^1$Ori C to similar HETGS spectra    
$\theta^1$Ori A and $\theta^1$Ori E. In both cases the spectra appear quite similar and require only three temperature
components between 5$\times10^6$ K and 3$\times10^7$ K, which may be a consequence of the much
lower statistics in these spectra. Again, the dominating contributions to the spectrum are in the
temperature components well above $10^7$ K. Such high temperatures are consistent with spectral
fits obtained form the zeroth order CCD spectra (Schulz et al. 2001). $\theta^1$Ori E shows Doppler velocities in the
lines of $\sim 500$ km/s (Ne X) significantly higher
line broading from the one observed in $\theta^1$Ori C. The lines in $\theta^1$Ori A however are
clearly broader with velocities of 900 km/s at Ne X.     

Most remarkable so far is the fact, that in both cases the models do not fit the spectra well unless
significant departures from the solar abundance distribution are introduced. Figure 2 shows preliminary  
model fits for $\theta^1$Ori A (top) and $\theta^1$Ori E (bottom). For clarity we only show the model and some of the brighter lines.
A full assessment of these spectra will be given in Schulz et al. 2002, in prep.). 
In both cases the spectra are dominated by strong Ne and and O lines,
whereas ionization states of Mg, Si, S, and Fe appear relatively weak. The high temperatures of the dominating components are
also determined by the underlying continuum, especially by the shape at short wavelengths. 
Given the similarity of the temperatures to $\theta^1$Ori C
we would not expect large deviations in the line emissivity from what is observed in Figure 1. 
For such high temperatures one would expect much higher emissivities at Fe XXIII and Fe XXIV 
(between 10 and 12 \AA~) and Fe XXV (1.85 \AA~). From the fit we determine an abundance for Fe relative
to solar of 0.2. Although also slightly underabundant relative to solar, Ne appears overabundant relative to almost all
the other elements, except Ni. The Ni XIX and Ni XXIV lines observed in $\theta^1$Ori A required an 
overabundance by about a factor 3.

\section{Discussion}

Given the assumption that the OB stars in the Orion Trapezium form a canonical sample with respect to
age, spectral type, and chemisty, we find remarkable differences in their X-ray spectra. 
In fact, the only similarities we find so far are multiple high temperature components 
and relatively narrow, symmetric, and unshifted lines. The temperatures in $\theta^1$Ori A and $\theta^1$Ori E
are not quite as hot as in $\theta^1$Ori C, but still high enough to question the validity of instability-driven
wind shock models. The fits do include temperature components with temperatures low enough to be consistent
with these models, however the contribution to the total spectrum is less than 20$\%$.
Because of the high temperatures Schulz et al. (2001) suggested the inclusion of magnetic field effects as well as
stellar rotation into the models for very young stars. Similarily,
Gagne et al. (2001) recently compared X-ray lines $\theta^1$Ori C with lines from another young
massive star $\tau$ Sco (B0 V, Cohen, Cassinelli, $\&$ Waldron 1997) and also concluded that these
properties may be the result interactions with magnetic fields.  

Most striking are clearly the observed abundance variations in $\theta^1$Ori A and $\theta^1$Ori E. Given the
fact that $\theta^1$Ori C does not require any awkward abundances and that those stars
evolved from a common progenitor cloud, such variations seem quite mysterious. However, one should
be quite careful with the interpretation of these findings, because due to the complexity of the spectra,
which are already reflected in the fact that we need multi-temperature components, many other issues can contribute.
One issue is possible contributions from low-mass companions. We know from recent interferometric
measurements that $\theta^1$Ori C (Weigelt et al. 1999) and $\theta^1$Ori A (Petr et al. 1998) have close low-mass
companions. Kastner et al. (2001) very recently reported on an iron deficiency, but neon enhancement in the
nearby classical T Tauri star TW Hya. With a luminosity of $\sim 10^{30}$ erg/s, however, a possible contribution
of a T Tauri to these spectra cannot exceed 10$\%$. So far $\theta^1$Ori E has not shown signatures of a companion.
Other issues involve opacity effects  - we did not yet introduce intrinsic absorption to each spectral
component - and wind interactions. An effect from the latter may not be unlikely in a close system like
the Trapezium, which also shows evidence for a variety of deeply embedded low-mass objects.
(Bally et al. 1998, Schulz et al. 2001).


\begin{references}
Bally, J., Sutherland, R.S., Devine, D., and Johnstone, D., 1998, \apj, 116, 293 \\
Bergh\"ofer, T.W., Schmitt, J.H.M.M., \& Cassinelli, J.P., 1996, \aap Suppl., 118, 481 \\
Chlebowski, T., Harnden, F.R. jr., and Sciortino, S., 1989, \apj, 341, 427 \\
Cohen, D.H., Cooper, R,G,m MacFarlane, J.J., Owocki, S.P., Cassinelli, J.P., \& Wang, P., 1996, \apj, 460, 506 \\
Cohen, D.H., Cassinelli, J.P, and Waldron, W.L., 1997, \apj, 488, 397 \\
den Herder, J.W., Brinkman, A.C., Kahn, S.M., et al., 2001, \aap, 365, L7 \\
Feldmeier, A., Puls, J., Reile, C., Pauldrach, A., Kudritzki, R.P., \& Owocki, S.P., 1995,  Ap$\&$SS, 233, 293 \\
Feldmeier, A., Kudritzki, R.P., Palsa, R., Pauldrach, A.W., \& Puls, J., 1997, \aap, 320, 899 \\
Gagne, M., Cohen, D., Owocki, S., and Ud-Doula, A., 2001, in X-ray at Sharp Focus, ASP Conference Series, eds. S. Vrtilek,\\
E.M.Schlegel, L. Kuhi \\
Hillenbrand, L.A., 1997, \aj, 113, 1733 \\
Hillier, D.J., Kudritzki, R.P., Pauldrach, A.W., Baade, D., Casinelli, J.P., Puls, J., and Schmitt J.H.M.M.,1993, \aap, 276, 117 \\
Houck, J.C., \& DeNicola, L.A., 2000, in Astronomical Data Analysis Software and Systems IX, ASP Conf. Ser.., Vol. 216, p.591 \\
Kahn, S.M., Leutenegger, M., Cottam, J., Rauw, G., Vreux, J.M., den Boggende, T., Mewe, R., \& Guedel, M., 2001, \aap, 365, L312 \\
Katama, Y., Koyama, K., Tsuboi, Y.,, and Yamauchi, S., 1997, \pasj, 69, 461 \\
Kastner, J.H., Huenemoerder, D.P., Schulz, N.S., Canizares, C.R, \& Weintraub, D.A., 2001, \apj, in press \\
Kudritzki, R.P., 1984, in ESA Fourth European IUE Conf., p 33 \\
Kudritzki, R.P., 1992, \aap, 266, 395 \\
Lamers, H.J.G.L.M., Haser, S., De Koter, A, \& Leitherer, C., 1999, \apj, 516, 872 \\
Lucy, L.B., \& White, R.L., 1980, \apj, 241, 300 \\
McFarlane, J.J., \& Cassinelli, J.P., 1989, \apj, 347, 1090 \\
Nagase, F., Ozawa, H., Ueda, Y., Ishida, M., and Dotani, T., 1999, in Star Formation 1999, Proceedings of Star Formation 1999,\\
ed. T. Nakamoto, 316 \\
Owocki, S.P., Castor, J.I., \& Rybicki, G.B., 1988, \apj, 335, 914 \\
Pauldrach, A.W., 1987, \aap, 183, 295 \\
Pauldrach, A.W., Kudritzki, R.P., Puls, J., Butler, K., \& Hunsinger, J., 1994. \aap, 283, 525 \\
Petr, M.G., Coud\'e du Foresto, V., Beckwidth, S.V.W., Richichi, A., and McCaughrean, M.J., 1998, \apj 500, 825 \\
Schulz, N.S., Canizares, C., Huenemoerder, D., Kastner, J.H., Taylor, S.C., \& Bergstrom, E.J., 2001, \apj, 549, 441 \\
Schulz, N.S., Canizares, C., Huenemoerder, \& Lee, J.C., 2000, \apj, 545, L135 \\
Seward, F.D, Froman, W.R., Giaconni, R., Griffith, R.E., Harnden, F.R. jr., Jones, C., Pye, J.P., 1979, \apj~ 234, 55 \\
Smith, R.K., Brickhouse, N.S., Liedahl, D.A., \& Raymond, J.C., 2001, \apj, 556, L91 \\
Tsuboi, Y., Koyama, K., Sakano, M., and Petre, R., 1997, \pasj, 49, 85 \\
Waldron, W.L., \& Cassinelli, J.P., 2001, \apj, 548, L45 \\
Weigelt, G., Balega, Y., Preibisch, T., Schertl, D., Sch\"oller, M., and Zinnecker, H., 1999, \aa 347, L15  \\
Yamauchi, S., Koyama, K., Sakano, M., and Okada, K., 1996, \pasj, 48, 719 \\

\end{references}
\end{document}